\documentclass[twocolumn, aps, pra, 10pt]{revtex4-1}
\usepackage{txfonts}
\usepackage{graphicx}
\begin{document}
\title{Intermittent optical frequency measurements to reduce the dead time uncertainty of frequency link}
\author{Hidekazu Hachisu} \email{Corresponding author: hachisu@nict.go.jp}
\author{Tetsuya Ido}
\affiliation{National Institute of Information and Communications Technology, Koganei, Tokyo 184-8795, Japan }
\begin{abstract}
The absolute frequency of the $^{87}{\rm Sr}$ lattice clock transition was evaluated with an uncertainty of $1.1\times 10^{-15}$ using a frequency link to the international atomic time (TAI). The frequency uncertainty of a hydrogen maser used as a transfer oscillator was reduced by homogeneously distributed intermittent measurement over a five-day grid of TAI. Three sets of four or five days measurements as well as systematic uncertainty of the clock at $8.6\times 10^{-17}$ have resulted in an absolute frequency of $^{87}{\rm Sr}\ {}^1S_0 - {}^3P_0$ clock transition to be 429 228 004 229 872.85 (47) Hz.
\end{abstract}

\maketitle

\section{Introduction}
Recent rapid progress of optical clocks has inspired various ideas of applications, attracting scientists outside the community of time and frequency standards. The community of geodesy is interested in optical clocks as detectors 
 which are sensitive to a difference of gravitational potentials\cite{geodesy}. Scientists propose to hunt for topological dark matter using clocks \cite{topology}. Network of atomic clocks combined with a quantum communication technique may realize secure communication in global scale \cite{quantum}. Realization of these applications practically depends on a frequency reference and a precision link to that reference. 
 This is particulary the case when an optical clock behaves as a sensing device by measuring frequency difference from another reference.

 Global navigation satellite system (GNSS) allows us to obtain the International Atomic Time (TAI)\cite{TAI} even at isolated locations. While TAI is normally used as a time scale, we can utilize it as a frequency reference of $\sim 10^{-16}$ accuracy because the frequency of TAI is, in most of the time, maintained according to the calibration served by cesium primary frequency standards. 
 Thus, commercial hydrogen masers (H-masers) with a satellite-based link to TAI, which we call ``TAI link'' hereafter, have often been employed in evaluating optical frequencies \cite{TakaJPSJ, Kohno, NRC, nict, KRISS, Wuhan, NMIJ} particularly in cases that the clock under test is developed or operated in laboratories where highly accurate primary frequency standards are not available.

The uncertainty of optical frequency measurement based on the TAI link has been limited at low $10^{-15}$ level predominantly owing to the link uncertainty. While it is expected that the link uncertainty is reduced by the integration of signals, the measurements have not been extensive. In addition, the least uncertainties of absolute frequency measurements were at the $10^{-15}$ level for long time, which hampered the rigorous validation of the frequency link at the $10^{-16}$ level. Following the advances of both optical clocks and cesium primary frequency standards in this decade, however, the latest absolute frequency measurements \cite{YbIonPTB, SYRTE, PTB, NPL} have uncertainties of the low $10^{-16}$ level, which provides environments to confirm the validity of the TAI link.

In this work, we extensively measured the absolute frequency of a $^{87}{\rm Sr}$ lattice clock using the TAI link. Fourteen days of frequency measurements, $10000 - 24000$ s per day, together with the compensation of the linear drift of a local transfer oscillator have reduced the link uncertainty to TAI down to $6.9\times 10^{-16}$. The resultant absolute frequency was consistent with other measurements performed in various institutes.

\section{Experiment}
Lattice clock of fermionic strontium is one of the most popular optical frequency standards. More than five institutes have $^{87}{\rm Sr}$ lattice clocks in operation to the best of our knowledge \cite{nict, SYRTE, PTB, NMIJ, Riken, JILA}. The apparatus of the lattice clock at NICT was briefly described in \cite{nict} when our clock based on a vertically oriented one-dimensional lattice started its operation.  Various parts of the setup have been improved since then to operate more stably and to reduce systematic uncertainties. The intensity of the Zeeman slower beam at 461 nm was increased more than twofold, allowing more atoms to be stably loaded to the lattice. The second stage cooling laser at 689 nm is now stabilized to a stable cavity made of ultralow expansion (ULE) glass. A commercial ytterbium (Yb) fiber comb which is stabilized to a H-maser was employed for the counting of optical frequencies. It is easier to achieve stable operation of long measurement time comparing with a Ti-sapphire based frequency comb that we used previously. A Ti-sapphire laser for the optical lattice is phase-stabilized to the nearest comb component. The clock laser is stabilized to a conventional cylindrical cavity of 10 cm length, which sits on a passive vibration isolation (PVI) platform. The spacer and mirror substrates are made of ULE glass. While the simple cylindrical shape is susceptible to vertical vibration, the vertical vibration is detected by an accelerometer, and the corresponding frequency shift is substantially compensated by a feed-forward method. The spectrum of vibration shows a peak at the mechanical resonance of the PVI platform. The signal of this resonant component detected by the accelerometer is digitally band-pass filtered and the corresponding frequency shift is applied to an acousto optic modulator, through which the clock laser is delivered to atoms and the frequency comb. The short term instability was evaluated by beating against another clock laser for a ${\rm Ca}^+$ ion clock \cite{Matsubara} where the vibration noise is suppressed in the same manner, resulting in a 1 s instability of $1.7\times 10^{-15}$ per clock laser. Note that a more elaborate feed-forward method was reported in \cite{NIST_ff}.

Thanks to these improvements, a Fourier-limited spectral width of 10 Hz is stably obtained using a probe pulse width of 80 ms, which we employed for an alternative atomic servo to evaluate various systematic shifts. The stability of the frequency measurement, on the other hand, is limited by the reference H-maser. Therefore, by reducing the probe pulse width to 50 ms, the measurement stability was not degraded but the digital stabilization to the atomic resonance was further secured. Frequency measurement was performed by counting the frequency with reference to a H-maser, HM4. The beat note between the 698 nm clock laser and the nearest comb component was sent to a zero dead time counter. The repetition frequency of the Yb comb was stabilized to HM4. The HM4 frequency is evaluated by a dual mixing time difference (DMTD) \cite{nakagawa}, which is a part of the Japan Standard Time (JST) system. On every second, the DMTD records the time difference of all H-masers and UTC(NICT), where UTC(NICT) is the local realization of Universal Coordinated Time (UTC) at NICT. 
Note that the difference of UTC and TAI is just cumulated leap seconds. In other words, both have an identical frequency.

\section{Systematic frequency shifts and uncertainties of Sr system}
Table \ref{tbl:sys} shows the systematic frequency shifts and uncertainties of the Sr system. Blackbody radiation causes the most significant frequency uncertainty similarly to other $^{87}{\rm Sr}$ lattice clocks operated at room temperature. Details of the systematic shifts are discussed as follows.

\begin{center}
\begin{table}[tbh]
\caption{Systematic shifts and uncertainties of the $^{87}{\rm Sr}$ lattice clock}
\begin{tabular}{lrr} \hline
Effect & Shift & Uncertainty \\
& $(10^{-17})$ & $ (10^{-17})$ \\ \hline
Blackbody radiation & -501.2 & 5.2 \\
Lattice scalar/tensor & 14.4 & 3.9 \\
Lattice hyperpolarizability & 0.2 & 0.1 \\
Lattice E2/M1 & 0 & 0.5 \\
Probe light & -0.2 & 0.1 \\
dc Stark & -1.0 & 4.7 \\
Quadratic Zeeman & -52.2 & 0.5 \\
Density & -2.5 & 2.5 \\
Line pulling & 0 & 0.6 \\
Servo error & -1.0 & 1.6 \\ \hline
Total & -543.5 & 8.6
\end{tabular}
\label{tbl:sys}
\end{table}
\end{center}

\subsection{Blackbody radiation (BBR) shift}
The vacuum chamber is made of aluminum alloy ($> 20$ mm thickness), which is more heat-conductive than stainless steel of $\sim$ mm thickness. Three thermistors with calibrations are distributed around the chamber to monitor the temperature. While the temperature of the laboratory $T_{\rm lab}$ is stabilized to $24 \pm 1 \ {}^\circ{\rm C}$, the temperature of the chamber $T_{\rm chamber}$ was affected by the cooling water of the Zeeman slower. Thus, the temperature difference between the chamber and the laboratory, which was no larger than $5 \ {}^\circ{\rm C}$ during the campaigns over more than two months, may cause a temperature gradient particularly on the glass windows. The maximum of the temperature differences among the thermistors was $0.49\ {}^\circ{\rm C}$, which decreased as $T_{\rm chamber}$ became closer to $T_{\rm lab}$.
 Atoms face two parallel windows made of BK7 glass of 15 mm thickness. The windows have open aperture of 45 mm diameter and are 23 mm apart from atoms. The outer edge of 22.5 mm width on the inner surface is behind the vacuum chamber from atoms with thermal contact. Considering these issues, we assumed that the temperature of the window center is the midpoint between $T_{\rm chamber}$ and $T_{\rm lab}$ and that the uncertainty of the window center is $(T_{\rm lab} - T_{\rm chamber})/2$. Furthermore, the temperature of the window surface is assumed to have linear dependence on the radial direction.
 In terms of the radiation from the oven, a mechanical shutter blocks the direct radiation to the atoms when the atomic transition is probed by the clock laser. 
 These considerations have resulted in the BBR shift of $ -501.2(5.2)\times 10^{-17}$. Note that the shift coefficient of BBR was measured accurately \cite{BBR1} and validated theoretically \cite{BBR2} with a corresponding uncertainty of less than $5\times 10^{-18}$, which is negligible at this stage.

\subsection{Lattice Stark shift}
A Ti:sapphire laser was employed for the optical lattice. The intensity was stabilized by monitoring the leak intensity behind the retro-reflecting mirror. The $1/e^2$ radii of the lattice beam and probe beam are 30 and 120 $\mu{\rm m}$, respectively. The depth of the lattice potential is $39 E_R$, where $E_R$ is the recoil energy of a lattice photon. The frequency is phase-stabilized to an Yb fiber comb, which is used for frequency counting. The polarizations of the lattice laser and the bias magnetic field are parallel. We interrogated $\pi$ transitions $(m_F = \pm 9/2 \rightarrow \pm9/2)$. Since this configuration is same as other works \cite{PTB, JILA}, we stabilized the lattice laser frequency to $f_{\rm lattice} = 368\ 554\ 465 \ {\rm MHz}$. This frequency was experimentally obtained as the magic frequency in \cite{PTB}. To characterize the lattice ac Stark shift, we performed two independent atomic servos of different lattice laser intensities at the laser frequencies of $f_{\rm lattice}$, $f_{\rm lattice} \pm 500\ {\rm MHz}$, and $f_{\rm lattice} \pm 1\ {\rm GHz}$. Linear fitting led to the magic point of the lattice laser frequency 368 554 527 (17) MHz. This frequency is higher than those in other studies \cite{PTB, JILA}. We suspect that this is attributed to the residual angle of lattice polarization and the bias magnetic field. The determination of the magic point free from the residual angle is beyond the scope of this work. The ac Stark shift at $f_{\rm lattice}$ was evaluated to be $1.44(39)\times 10^{-16}$. Note that the other effects due to the hyperpolarizability and E2-M1 coupling\cite{Lemonde} are not significant in the current stage of measurement. They are all at the low $10^{-18}$ level.

\subsection{Zeeman shift}
The correction signal of the atomic servo was obtained from the average frequency of two $\pi$ transitions from $m_F = \pm9/2$ states. Thus, the first-order Zeeman shift is not relevant. The correction due to the quadratic Zeeman shift is determined using the magnitude of the bias magnetic field, which was evaluated from the Zeeman splitting of two transitions. The splitting is typically 951 Hz, which corresponds to a quadratic Zeeman shift of $−5.2\times 10^{-16}$. We employed the shift coefficient obtained from the weighted average of five other measurements \cite{JILA, Ludlow, Lemonde, Falke, Bloom}. The fractional uncertainty is $5\times 10^{-18}$ due to the uncertainty of the coefficient.

\subsection{Density shift}
The density shift was evaluated using an alternative atomic servo of more than 50000 s with two different atom densities. The result indicated a density shift of $ - 2.5(2.5)\times 10^{-17}$.

\subsection{dc Stark shift}
The vacuum chamber has two large glass windows as described above. The dc Stark shift due to stray electric charge on these windows was investigated according to the method first proposed in \cite{SYRTEdc}. An external electric field was imposed on the atoms from transparent electrodes placed outside the chamber. The electrode on a round substrate is divided into three parts. Two minor electrodes are of rectangular shape and located at the vertical and horizontal edge of the substrate, enabling an electric field with small components parallel to the window. External fields, (i) perpendicular to the windows, (ii) horizontally tilted, and (iii) vertically tilted are added and the dc Stark shifts in these cases are measured by an alternative operation with or without the electric field. By applying a voltage of $\pm 500$ V at maximum to the two electrodes, we observed a sub-Hz dc Stark shift. Quadratic fitting indicated that the component of the stray electric field parallel to the external field causes the frequency shifts of (i) 1.0, (ii) 2.2, and (iii) 1.4 mHz. The atoms are loaded horizontally from an atom oven. Thus, the system is asymmetric in the horizontal direction. The rather large stray electric field in (ii) is consistent with this asymmetry. In addition, the coefficient of the quadrature component is similar in (ii) and (iii), confirming the symmetry of additional electric field. By decomposing the shift to three orthogonal directions, the corresponding shifts of each component are 1.0 mHz (perpendicular to the window), 2.6 mHz (horizontally), and 0.6 mHz (vertically). These results and standard errors of the parabola fitting estimates that the total dc Stark shift is $(−4.2\pm 20)$ mHz.

\subsection{Probe Stark shift}
The clock laser slightly couples the ground and excited states to other electronic states, causing an ac Stark shift. 
  We adjusted the probe intensity so that the probe pulse width of 50 ms corresponds to the $\pi$ pulse of the Rabi oscillation. This pulse width, the transition strength theoretically predicted in \cite{Santra}, and the shift coefficient obtained in \cite{Baillard} lead to the probe light shift of $ −2(1)\times 10^{-18}$.

\subsection{Other shifts and total uncertainty of Sr system}
There are other minor shifts such as the second-order Doppler shift. However, their contributions are negligible and have an uncertainty level of $10^{-18}$. The uncertainties described above are not correlated. Thus, the square root of the sum of the squares results in a total uncertainty of $8.6\times 10^{-17}$.

\section{Data analysis}
\begin{figure}[tbh]
\centerline{\includegraphics[width=\columnwidth]{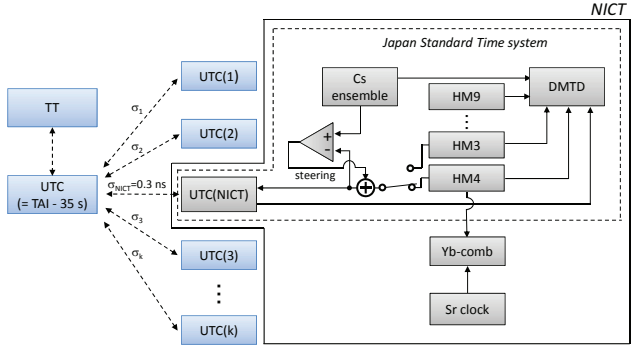}}
\caption{Scheme of the frequency link between the Sr lattice clock and TAI. The frequencies of UTC and TAI are identical. The time difference of 35 s is the cumulated leap seconds. HM4 was the source oscillator of UTC(NICT) during the measurement. The time link between UTC and UTC(NICT) has a type-A uncertainty of 0.3 ns, which is a major uncertainty of TAI-based frequency measurement. The scale interval of the TT (Terrestrial time) is the SI second on the geoid. See text for other parts of the scheme.}
\label{Schematic}
\end{figure}

We performed the frequency measurement with reference to a H-maser. Since the inaccuracy of the Sr system is much less than the fluctuation of the H-maser, the uncertainty of the absolute frequency measurement is determined by the evaluation of the H-maser frequency. The overall scheme of this evaluation is depicted in Fig. \ref{Schematic}. We employed TAI as a reference for the evaluation. TAI is based on the average of more than 400 commercial atomic clocks that are operated in various institutes worldwide. Participating institutes send two kinds of data to the Bureau International des Poids et Mesures (BIPM). One is the time differences of every fifth day between their commercial atomic clocks and UTC(k), where UTC(k) is the local realization of Universal Coordinated Time at laboratory ``k''. The other data is the output of a GNSS receiver or the result of two way satellite time and frequency transfer. These systems are operated in respective institute “k” with its local clock obtained from UTC(k). Thus, the data effectively connects the clocks in various institutes. BIPM calculates the weighted-average ``tick'' of commercial atomic clocks on every fifth day and adds a correction to the average so that the frequency is consistent with the SI second. The SI second is determined from the weighted mean of primary frequency standards (PFSs) or secondary frequency standards operated in national metrological institutes through the differential frequency of their time scale against their own PFSs. BIPM reports the result of the calculation every month in the Circular T as the time difference of UTC and UTC (k) on every five days.

Infrequent ticks of TAI cause a difficulty in utilizing TAI as a frequency reference. The clock rate can be compared only for the frequency of a five-day average. In principle, we need to continuously operate clocks under test for five days, which is still not easy for lattice clocks. Thus, for the frequency evaluation based on TAI, we need a stable local transfer oscillator (LTO) that oscillates continuously for five days. The most commonly used device for this purpose is a H-maser, which we also employed in this work. Another possibility could be a single-ion clock that stably operates for a long time. The absolute frequency of a Sr lattice clock at PTB \cite{PTB} was measured using an ${\rm Yb}^+$ ion clock \cite{YbIonPTB} as a transfer oscillator that connects the Sr lattice clock and Cs primary frequency standards in the same campus.

Figure 2 depicts the basic scheme of absolute frequency measurement using TAI. Here, the horizontal and vertical axes are time and fractional frequency against TAI, respectively. $y_{\rm UTC(k)}$ represents the temporal frequency variation of UTC(k). The area A, which is the integrated fractional frequency difference between UTC(k) and TAI, is the variation of time difference between 0:00 of day 0 and that of day 5. In other words, the area A is expressed using the following formula

\begin{eqnarray*}
A & = & \int_{Day0}^{Day5} \left( y_{\rm UTC(k)}-1\right) dt\\
& = & {\rm \left[ UTC(k)-UTC\right]_{Day5} - \left[ UTC(k)-UTC \right]_{Day0}}.
\end{eqnarray*}

Here, the time difference $\rm [UTC(k)-UTC]_{Day0, Day5}$ is available in Circular T. The operations of optical clocks, depicted as a bar in Fig. 2(a), are often  limited in short time, which requires adding uncertainty to estimate the frequency of UTC(k) at that instance of measurement. We call this uncertainty dead-time uncertainty in this work because the five days have intervals when the lattice clock is not in operation.
 Using a more stable oscillator as an LTO reduces this error. Since the LTO oscillates continuously for five days, the sum of areas A and B in Fig. 2(b) can be obtained as

\begin{eqnarray*}
&    & A+B \\
& = & \int^{\rm Day5}_{\rm Day0}\left( y_{\rm LTO}-1\right) dt \\
& = & \int^{\rm Day5}_{\rm Day0}\left( y_{\rm UTC(k)}-1\right) dt + \int^{\rm Day5}_{\rm Day0}\left( y_{\rm LTO}-y_{\rm UTC(k)}\right) dt\\
& = & \left\{ \left[\rm UTC(k)-UTC\right]_{\rm Day5}-\left[\rm UTC(k)-UTC\right]_{\rm Day0}\right\} \\
&   & +\left\{ \left[\rm LTO - UTC(k)\right]_{\rm Day5}- \left[\rm LTO - UTC(k)\right]_{\rm Day0}\right\} \\
& = & \left[ \rm LTO-UTC\right]_{\rm Day5} - \left[ \rm LTO-UTC\right]_{\rm Day0}.
\end{eqnarray*}
Here, frequency measurement is affected not by the fluctuation of UTC(k) but by LTO. The variation of time difference [LTO - UTC(k)]  over five days is locally obtained by a time interval counter or DMTD system.

\begin{figure}[tbh]
\centerline{\includegraphics[width=\columnwidth]{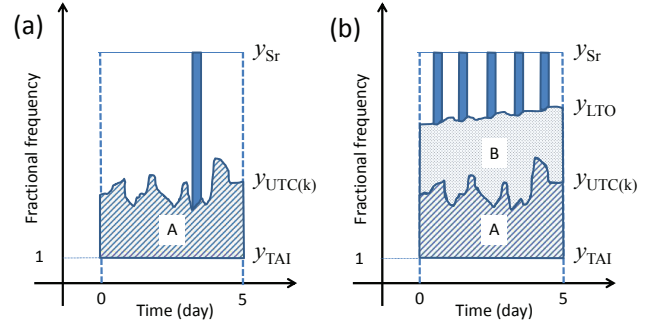}}
\caption{Scheme for measuring absolute frequency using TAI. (a) Circular T issued by BIPM provides the time difference between UTC(k) and UTC at 0:00 (UTC) of day 0 and day 5, which corresponds to area A. In the case of UTC(NICT), the frequency is sporadically adjusted for the $10^{-15}$ level with reference to the Cs ensemble. Thus, the frequency of UTC(NICT) at the moment of measurement is not well known, causing an error in the frequency link. (b) A highly stable local transfer oscillator (LTO) reduces the ambiguity of LTO frequency. Area B is attained by a time interval counter.}
\label{fig2}
\end{figure}

We normally used HM4 as the LTO. HM4 is the most stable in the H-maser array of JST. The instability is shown in Fig. 3 as the blue curve. Frequency measurements of our lattice clock with reference to HM4 were performed for several hours per day for five (or four) consecutive days. Since the frequency measurements fit the five-day grid of TAI, we can obtain the five-day average of the maser frequency with reference to TAI.  Figure 4(a) shows a result of the clock frequency measurement on MJD $57124 - 57128$. The systematic shifts discussed in Sec. 3 are already corrected according to the error budgets of the respective day. Since the inaccuracy of the Sr lattice clock is less than $1\times 10^{-16}$, the overall slope indicates the linear drift of the HM4 frequency. The measurement interval of each point determines the magnitude of the error bar according to the instability shown in Fig. 3. The linear drift of the LTO causes errors, particularly for the measurement on day 1 or 5. In other words, symmetrically balanced weighting of the measurements over five days reduces this error. Considering these issues, we performed the measurement every day for similar durations of $10000 −- 24000$ s. Furthermore, we obtained the linear drift rate of the H-maser with reference to strontium, and subtracted this part of drift from the beat frequency. This mitigates the residual asymmetry of measurements over five days. The result of compensating the H-maser drift as well as incorporating the five-day average of HM4 relative to TAI is shown in Fig. 4(b) as filled circles. The residual fluctuations are mainly due to the fluctuation of the LTO, which was confirmed by changing the LTO to another H-maser, HM9. The instability of HM9 is shown as the green curve in Fig. 3, indicating an excess long-term instability compared with HM4. We can obtain the frequencies based on HM9 from same data of optical beat frequency because the relative frequency of HM9 against HM4 is always monitored by the DMTD system. We found that the standard error of the measurement with reference to HM9 was 2.8 times larger than that with reference to HM4.

\begin{figure}[htb]
\centerline{\includegraphics[width=\columnwidth]{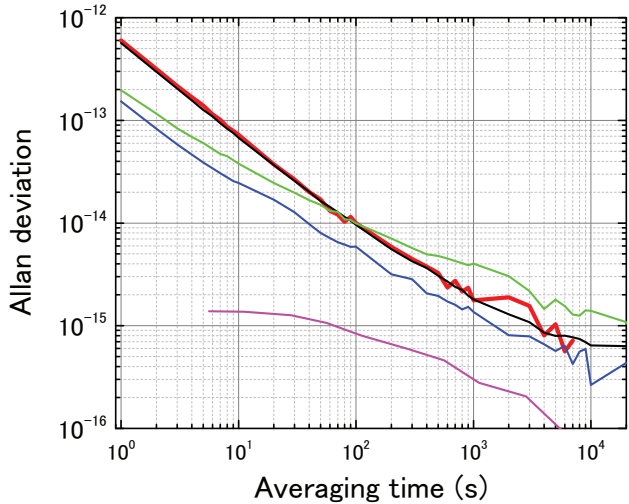}}
\caption{Allan standard deviations relevant to this work. The instability between the Sr lattice clock and the Yb comb referenced to HM4 is shown as a red curve. Blue and green curves show the instabilities of H-masers HM4 and HM9, respectively. They are calculated by extended three-corner-hat method using several H-masers’ data in NICT. The bandwidth of the phase noise measured by JST system (blue and green) is limited to 10 Hz. The transfer of the maser signal to laboratory might be another reason why red is not as stable as blue. Thin black curve is the instability of simulated oscillator which was used to estimate the dead time uncertainty (see text). The instability of an interleaved stabilization of the Sr frequency standard is drawn as a purple curve. }
\label{fig3}
\end{figure}

\begin{figure}[htb]
\centerline{\includegraphics[width= \columnwidth]{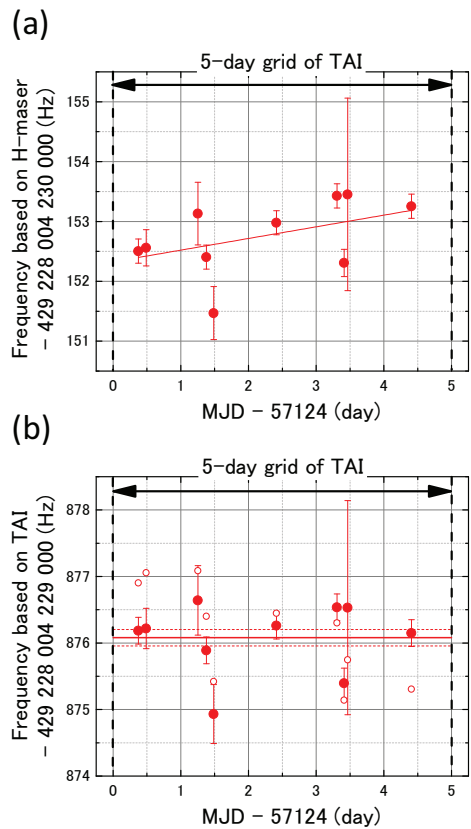}}
\caption{(a) A result of 5-day frequency evaluation with reference to a H-maser (HM4). The correction of the systematic shifts were made on each day. The overall slope indicates the frequency drift of H-maser. (b) Filled circles are frequencies based on TAI derived from same raw data as (a). The frequency drift is eliminated according to the drift rate obtained in (a). The H-maser frequency is calibrated by the TAI link. Empty circles are obtained with a single value of HM4 calibration, which was derived from  the five-day average of UTC(NICT) frequency. The drift of the UTC(NICT) is clearly observed. }
\label{fig4}
\end{figure}

In most of the previous measurements using TAI, the absolute frequencies were evaluated using the scheme shown in Fig. 2(a). Since no care was taken for the distribution of the measurement time over the 5-day grid of the TAI, the measurement suffered from large dead time uncertainty of  $> 2\times 10^{-15}$. The frequency evaluated without drift compensation is depicted as empty circles in Fig. 4(b). UTC(NICT) adjusts the frequency every 8 hours referring to ensemble of Cs atomic clocks. Thus, sporadic frequency changes may occur, which required a large dead time uncertainty of $2.6\times 10^{-15}$ in previous work \cite{nict}. Nevertheless, the difference between the two weighted means obtained from the 5-day measurement was only 0.10 Hz $(2.3\times 10^{-16})$, suggesting that the five-day consecutive measurement reduces the dead time error.

The dead time uncertainty was estimated by numerical simulation as first investigated in \cite{Yu}, where the advantage of the distributed dead time was described.
 The estimation of dead time uncertainty assumes negligible linear drift of the frequency reference. Maser ensemble and TAI were examples, where their linear drifts are reduced by referring the past signals. On the other hand, we canceled the linear drift of the H-maser by referring to the atomic resonance of the $^{87}{\rm Sr}$ clock transition. The removal of the H-maser drift shown in Fig. 4 effectively allows the use of a drift-compensated reference. Following the method described in \cite{Yu}, we first modeled the noise characteristics of frequency measurement as shown in Fig. 3 (black curve). The four simulated noises incorporated are $\sigma(\tau ) = 5\times 10^{-13} \tau^{-1}$ for white PM, flicker PM with $\sigma = 2\times 10^{-13}$ at $\tau = 1$ s, $3\times 10^{-14} \tau^{-1/2}$ for white FM, $3\times 10^{-16} \tau^0$ for flicker FM, and $5\times 10^{-18} \tau^{1/2}$ for random-walk FM, where $\tau$ is the integration time in seconds. Then, commercial software ``Stable 32'' generated the time series of the frequency for $86400\times 5$ s, and from which five of the 15000 s data were equidistantly extracted. It is known that the noise generation employed in the software is based on \cite{Kasdin} and \cite{Greenhall}.
 We generated ten time series of frequency. The deviations of the average of 15000 and $15000\times 5$ s from the total mean of $86400\times 5 $ s correspond to the dead time uncertainties of one-day and five-day measurements, which were $1.1\times 10^{-15}$ and $2.7\times 10^{-16}$, respectively. We also investigated the dead time uncertainty in the case that HM9 is employed for the LTO. To simulate the green curve in Fig. 3, the numerical model includes phase noise $\sigma(\tau)$ of $1.5\times 10^{-13} \tau^{-1/2}$ for white FM  and $1.5\times 10^{-15}$ for flicker FM. The simulation resulted in the dead time uncertainty of $2.1\times 10^{-15}$ and $7.0\times10^{-16}$ for the one-day and five-day measurements, indicating the significance of preparing an LTO with an appropriate stability.

Three sets of frequency measurements were performed. Figure 4 shown above is the result of the last campaign (Campaign \#3). Measurements were performed on four consecutive days for Campaign \#1, and the other campaigns \#2 and \#3 were performed on five days. The result is summarized in Fig. 5, where the uncertainties shown as thick error bars are statistical uncertainties derived from the standard deviation of the mean of the respective 5 (or 4)-day measurement. Note that the corrections to the Fig. 4(b) due to the gravitational red shift and the calibration of TAI to the SI second, have resulted in the absolute frequency represented in Fig. 5. The total weighted average is calculated on the basis of these three means with the statistical weighting. The weights of campaigns \#$1 - 3$ were 16.4, 65.8, and 17.8\%, reflecting the scattering of beat frequencies that presumably depends on the fluctuation of HM4 during the respective campaign. The thin error bars for the three campaigns indicate the total uncertainties including the systematic uncertainty of the frequency link. Table II shows all relevant uncertainties, where the left and right columns show the uncertainty of Campaign \#3 and that of the total weighted mean, respectively. The source of systematic uncertainty is divided into two parts, corresponding to the strontium atom system, and the frequency link. The statistical uncertainty is the standard deviation of the mean. Systematic uncertainty of the strontium system is described in Section 2. We need to consider the gravitational red shift for the frequency link to the SI second on the geoid. The geometrical height of the laboratory was measured in 2008. Despite the large earthquake in 2011, it is known that the elevation of western area of Tokyo changed less than 5 cm. Using the Japanese geoid model GSIGEO2011, the geoid height of the clock was determined to be 76.5(2) m, which requires correction of resonant frequency for $-8.32 \times 10^{-15}$. Considering the tidal effect, we introduced the uncertainty of the red shift equivalent to a geoid height of 1 m, which corresponds to $1.1\times 10^{-16}$. The uncertainty of time difference $\rm HM4 - UTC(NICT)$ is small. It corresponds to the frequency difference of five day interval $5\times 10^{-17}$ according to a zero measurement using a common signal introduced to DMTD. For the link between UTC(NICT) and TAI, the type A uncertainty of 0.3 ns is suggested in Circular T . Two time differences at the start and end of the five days determine average differential frequency of TAI and UTC(NICT) with an uncertainty according to Eqn. (25) in \cite{Panfilo}. The uncertainty of five day average is $( \sqrt{2}\times 0.3\ {\rm ns}) / (5\times 24\times 3600\  {\rm s}) = 9.8\times 10^{-16}$. Finally, the corrections and uncertainties of TAI relative to the SI second were considered. The frequency difference of TAI from the SI second is found in Circular T as a 30-day average, which again requires the evaluation of the dead time uncertainty because five day averages of TAI frequency are used for each campaign. We calculated this dead time uncertainty in same manner as HM4 dead time simulation described above. Here, the reference is the EAL (Echelle Atomique Libre, free atomic time scale), whose instability is found in Circular T as quadratic sum of three components with $\tau_d$ in days: (1)$1.7\times 10^{-15} \tau_d^{-1/2}$, (2) $3.5 \times 10^{-16} \tau_d^0$, (3) $2\times 10^{-17}\tau_d^{1/2}$. The calculation indicated that the dead time uncertainty of $f({\rm TAI})$ against the SI second is $1.1\times 10^{-15}$ for five day average.

\begin{figure}[tbh]
\centerline{\includegraphics[width=\columnwidth]{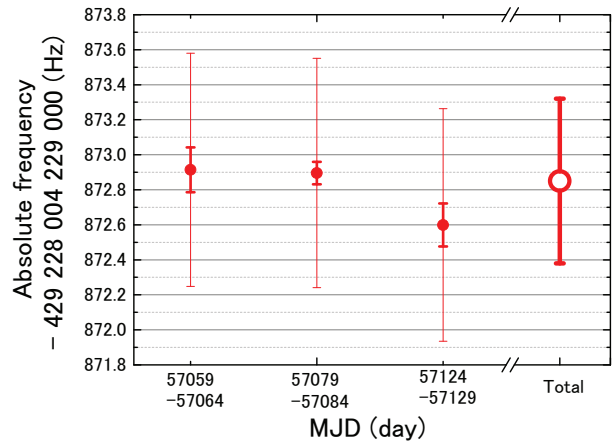}}
\caption{Absolute frequency of three sets of 5(or 4)-day average. Thick and thin error bars are statistical and systematic uncertainties. The later includes the uncertainty due to the frequency link to TAI. Total average (empty circle) was calculated according to the weights based on the statistical uncertainties.  }
\label{fig5}
\end{figure}

\renewcommand\thefootnote{\it \alph{footnote}}
\begin{table}[tbh]
\caption{Uncertainty budget of the frequency measurements.}
\begin{tabular}{lrr} \hline
Effect& Campaign \#3  & Total mean \\
& & of three campaigns \\
& $(10^{-17})$ & $ (10^{-17})$ \\ \hline
Statistical & 29 & 19 \\
{\bf Strontium} \\
Systematic & $9$\footnotemark & 10 \\
{\bf Link} & & \\
Gravity & 11 & 8 \\
Dead time (HM4) & 27 & 19 \\
HM4 - UTC(NICT)  & 5 & 4 \\
\hspace{5mm} (DMTD, 5days) & & \\
UTC(NICT) - TAI & 98 & 69 \\
Dead time (TAI) & 110 & 76 \\
$f(\rm TAI-TT)$ & 26\footnotemark[2]  &  25\footnotemark[3] \\ \hline
{\bf Total} & 155 & 110 \vspace{2mm}\\
$^a$ From Table I & & \\
$^b$ Circular T No. 328 & & \\
$^c$ Circular T No. 326 - 328 & &
\end{tabular}
\label{tbl:syslink}
\end{table}

The uncertainty of the weighted mean (right column) was determined as follows. In terms of the strontium atom system, the statistical uncertainty is the standard deviation of the mean of the three points in Fig. 5. The systematic uncertainty of strontium system is derived from weighted mean of three systematic tables as the systematic uncertainty of the BBR varies in three campaigns and those are not random. On the other hand, some biases due to the link have characteristics of a statistical nature. Thus, the uncertainty of the weighted mean would be less than that for one five-day evaluation. The bias due to the tidal effect was statistically averaged. According to Eqn. (25) in \cite{Panfilo}, the reduction of $\rm UTC - UTC(NICT)$ link uncertainty would be $1/(14({\rm day})/5({\rm day)})^{0.9} = 40\%$ if three sets of 5(or 4)-day units are seamless. However three campaigns are separated in time in this case. Therefore, three are assumed to be independent and we should adopt the factor $1/\sqrt{3} = 58\%$ for the case of three equal weights. The deviation from the equal weight mitigates this averaging effect to about 70\%. The total uncertainty of $f({\rm TAI-TT})$ is almost same as that of 5-day because it may include systematic bias of Cs fountains. Finally, the total uncertainty of our measurement was determined to be $1.1\times 10^{-15}$, and the absolute frequency of the$^{87}{\rm Sr}$ lattice clock transition was concluded to be 429 228 004 229 872.85 (47) Hz.

Figure 6 summarizes the record of absolute frequency measurements performed in various institutes worldwide. The result reported here is consistent with other measurements within the uncertainty, demonstrating the validity of the frequency measurement based on the TAI link.

\begin{figure}[tbh]
\centerline{\includegraphics[width=\columnwidth]{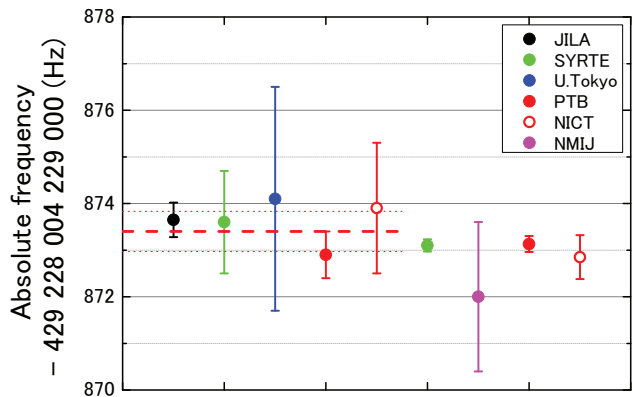}}
\caption{Record of absolute frequency measurements performed in various institutes. The frequency is found in publications: JILA \cite{Gretchen}, SYRTE \cite{EPJD, SYRTE} U. Tokyo \cite{Hong}, PTB \cite{Falke, PTB}, NICT \cite{nict}, NMIJ \cite{NMIJ}. The last data is the result of this work. Current CIPM recommendation and its uncertainty are expressed as broken lines, which were determined in 2012 based on the first five data.  }
\label{fig6}
\end{figure}

\section{Conclusions}
In summary, the absolute frequency of the $^{87} {\rm Sr} \ {}^1S_0 - \ ^3P_0$ clock transition was evaluated with an uncertainty of $1.1\times 10^{-15}$ using the TAI link. Three sets of five (or four)-day measurements, where the frequency is measured for $10000 -− 24000$ s per day, have reduced the uncertainty due to the frequency link between TAI and LTO. The uncertainty due to the measurement dead time is evaluated by numerical simulation. These efforts reduced the link uncertainty.

The intermittent comparison of an optical clock frequency can be applied to the estimation of the TAI frequency, similarly to the role played by microwave fountain standards.  Cesium or rubidium fountain frequency standards have so far contributed to the calibration of TAI frequency by their continuous operation over 15 days or more. Optical clocks combined with a stable LTO may not require continuous operation as they reach the same level of instability in a few hours. While frequency link in principle requires the prolonged signal integration, the homogeneously distributed intermittent operation of optical clocks as well as the management of dead time error could be sufficient for the estimation of the TAI frequency. Such measurements simultaneously provide an alternative of the frequency link for optical clocks. Link technique is always verified by a comparison with another independent method. The frequency evaluation investigated here may play a role in validating a novel link technique such as satellite-based transfer using two-way carrier phase \cite{miho, hachisu}. The reduction of the link uncertainty will enable certain applications of optical clocks that require the operation in the field or isolated locations.

The authors thank F. Nakagawa and H. Ito for providing the record of JST. We are also grateful to Y. Hanado, M. Kumagai, and T. Gotoh for the discussion regarding the link uncertainty. Parts of the Sr system were built by A. Yamaguchi, A. Nogami, S. Nagano, and Y. Li in the early stage. H. Ishijima, S. Ito, and M. Mizuno provided the necessary technical assistance for the measurements.

\end{document}